# Research on the evolution of domestic multifunctional meter technology

Zhen Zhang[1]

1. Huaneng Jinan Huangtai Power Generation Co., LTD., Jinan, Shandong 250100

Abstract: The technical evolution of domestic multi-functional electricity meter is deeply discussed. With the rapid development of the domestic power market and the continuous innovation of technology, the domestic multi-functional electricity meters have experienced the transformation from simple billing to complex multi-functional, from a single application to a wide range of fields. This transformation has not only driven the rapid development of electricity meter technology, but also met the increasing power demand and management requirements. This paper expounds the concept of multi-function meter, the working principle and algorithm of digital multiplier, the initiation and evolution of multi-function electricity meter standard, and the initiation and evolution of domestic multi-function electricity meter products. Although the domestic independent production of multi-functional meter has made great achievements in performance, but in the reliability and key process technology still need to be improved. In addition, the development of communication technology also provides a new opportunity for the progress of electricity meter technology. The application of the new technology provides a more convenient and efficient way for the data transmission and remote management of electricity meters. Domestic multi-functional electricity meters have made remarkable achievements in technology evolution and application and expansion, but they still face some challenges and opportunities. In the future, with the continuous development of the power market and the promotion of smart grid construction, domestic multi-functional electricity meters need to continue to strengthen technological innovation and product research and development, improve the reliability and competitiveness of products, in order to meet higher application needs and market requirements.

keyword：Multifunctional meter　analog multiplier　dataprocessing unit

Chinese chart classification number:TM933.4

about the author:Zhen Zhang (1977-) is engaged in research on electric metering technology 721047546@qq.com

# 0 Introduction

In the modern power system, multi-functional electricity meters play a vital role. It is not only the basic equipment of power system operation monitoring, but also an important tool to achieve the goals of energy management, economic accounting and energy conservation and emission reduction.

The multi-functional electricity meter can monitor and record the key parameters such as voltage, current, active power, reactive power and power factor in real time, providing accurate data support for the monitoring and management of the power system. Through the analysis of these data, the power operators can understand the operation status of the system, find and deal with potential problems in time, and ensure the stable operation of the power system. Multi-functional electricity meters can provide detailed electricity consumption data, and help power users and enterprises in energy management and economic accounting. Through the analysis of power consumption data, users can find out the waste points in energy use, and formulate energy-saving measures to reduce energy costs. At the same time, these data can also be used for the transaction and settlement of the electricity market, providing a strong guarantee for the fair and transparent operation of the electricity market. Multi-functional electricity meter provides data support for energy conservation and emission reduction of energy consumption and environmental protection of power system. Through the analysis of power consumption data, power operators can optimize the system operation mode, reduce emissions and pollution, and achieve a green and low-carbon power supply. Multi-functional electricity meters play a vital role in modern power systems. With the continuous progress of technology and the continuous improvement of the application demand, the multi-functional electricity meter will continue to develop and improve, providing a strong guarantee for the stable operation and sustainable development of the power system. At present, there is no unified international standard of multi-functional electricity meter in the world. In China, since 1992, the concept, standards and products of multi-functional electricity meters have been introduced, improved and tended to be complete.

## 1. Concept of multi-function table

This concept was first published in March 1995 in the second part of "Technical Report on Power metering Modernization-Power Measurement Technology of East China Power Grid": "Improve time-sharing metering technology, steadily develop multi-function meter".

1) Multi-function meter concept: The report states: " The development of electronic technology has created the foundation for multi-function electricity meters. In addition to time-sharing measurement, the multi-functional electricity meter generally also includes demand measurement and load control, load survey representative daily load curve and other functions, simplifying the user measurement point meter, convenient data collection, should pay attention to promote its development ".

2) The process of writing the report

- -In 1992, the Technical Committee of East China Electric Power Administration set up a research group to draft it. The research group is composed of 14 experts from the East China Electric Power Administration bureau and its provincial power grid.

- -In 1993, the subject materials were completed successively.

- -In 1994, Zeng Naihong of East China Electric Power Administration Bureau searched and wrote, printed the first draft and finally wrote the report.

3) Comments on this article

In the start and development of domestic multi-functional meters, the former Electric Power Department of the Ministry of Energy and the East China Electric Power Administration pave the way, and their achievements in promoting the development of domestic high-end electricity meters:

- -In China, issued the regional power grid power metering modernization technology report for the first time, and put forward the development and application of domestic multi-functional meters.

- -Organize the drafting of the first domestic multi-functional table mechanical standard and power

standard, to provide a basis for the development and application of domestic multi-functional table products.

## 2. The working principle and algorithm of electric energy meter using digital multiplier

The main content of this part is extracted from "Key Technology of Dynamic Power Measurement of New Power System" by China Electric Research Institute and edited.

1) The working principle of the electric energy meter using the digital multiplier

- -Metering method of active and active electric energy:

By (AC voltage sampling-analog-to-digital converter ADC circuit and AC current sampling-ADC-high pass filter HPF circuit) parallel digital multiplier-low pass filter LPF-correction-integral-digital frequency conversion-pulse output-active energy.

- -Metering method of reactive electric energy:

After parallel connection with active power metering (AC voltage sampling-AC DC circuit-ADC-HPF circuit), then connect to Hilbert filter-digital multiplier-LPF-correction-integration-digital frequency conversion-pulse output-reactive power.

2) The power algorithm of the digital multiplier

-Time domain: point product and numerical integration algorithm; Walsh overlapping algorithm; weighted average algorithm and window function convolution algorithm.

-Frequency domain: (window interpolation) DFT algorithm; FFT algorithm.

-Time and frequency domain: wavelet transform; S transform; modern spectral estimation and intelligent algorithm.

The above algorithm can calculate electrical parameters: active power / power; reactive power / power; dependent power / power; voltage and current effective value and base wave frequency.

- -Typical case of electric energy meter integration algorithm

This is an excerpt from the 2010 Weisheng "The application of Newton-Cotes Integral algorithm in electric energy metering".

The active power metering of DTSD341-MA1-type and 0.1S high-accuracy high-order integral algorithm is adopted.

For the signal containing the highest n harmonic, to accurately calculate the signal power, power, effective value and other parameters, the sampling of the n harmonic signal must be more than 3 points, that is, the sampling frequency of the signal must ensure that each base wave cycle signal sampling more than 3n points.

To improve the accuracy of dot product algorithm for any determined input signal, higher-order integration algorithm can be used. The Newton-Cotes algorithm is a practical and easily realized numerical quadrature algorithm. The Newton-Cotes algorithm of n order has at least n times algebraic accuracy.

## 3. The onset and evolution of multi-functional electricity meter standards

The technical requirements of electricity meter definition, electricity meter type, accuracy and multi-functional in the standard of multi-functional energy meter reflect the technical development level of domestic multi-functional meter in different periods.among:

The multifunction table is the most complex function, which can be divided into two categories: basic functions and extended functions.

Category 1 of basic functions. After the summary and analysis of 7 standards of multi-function meter, smart electricity meter and new generation smart electricity meter (excluding the standard of intelligent Internet electricity meter), among them, the state Grid standard Q / GDW 13542013 "Functional Specification of Smart Electricity Meter" has complete functions and forward-looking, and can be used as a reference level for comparing different standards. Its functions are divided into two categories: category 1 data / information / software output function, including general (multifunctional meter, smart meter) output function and special (smart meter) output function; category 2 internal data / information / software formation and processing function.

1) Reference level: In March 2013, the state Grid standard Q / GDW 13542013 "Functional Specification for Smart Electricity Meters" was released.

First, the definition of smart electricity meter: composed of measurement unit, data processing unit, communication unit, etc., with electrical energy measurement, information storage and processing, real-time monitoring, automatic control, information interaction and other functions.

Second, the type of electricity meter: the static electricity meter

Third, accuracy: electricity meters can be divided into four levels: 0.2S,0.5S, 1 and 2.

Fourth, function

-Basic functions of category 1 (30 item 119)

Type 1 data / information / software output function (2395)

The following is the general (multi-function meter, smart meter) output function (14 items and 58 detailed rules):

First, the measurement of one-way or two-way total active electric power and phase electric power, one-way or four-quadrant total reactive power-1 rules (do not use the phase active power to calculate the total active power).

Second, the metering combined active electric energy and the combined reactive electric energy.

Third, through software programming, the special combination of reactive electric energy.

Fourth, the measurement of active electric energy, reactive electric energy-2 detailed rules.

Fifth, the demand measurement-5 detailed rules.

Sixth, the clock and the school time-5 detailed rules.

Seven, the event record-22 detailed rules.

Its eight, communication requirements-8 detailed rules.

Nine, communication mode type-5 detailed rules (RS485, infrared communication, carrier communication, public network communication, micro-power wireless communication).

Its ten, signal output-3 detailed rules (pulse output, multi-function signal output, control output).

Its eleven, power grid operation parameters monitoring-4 detailed rules.

Twelve, security protection-3 rules (zero, programming, parameter setting)

Its thirteen, constant magnetic field monitoring

Its 14, electricity meter software comparison function

The following is the special (smart meter) output function (9 items and 37 detailed rules):

First, the cost control function-14 detailed rules.

Second, the load record-3 detailed rules.

Third, the ladder of electricity price-5 detailed rules.

Fourth, power failure meter reading-2 detailed rules.

Fifth, protect the electricity function-5 detailed rules.

Its six, the alarm-3 detailed rules.

Seven, auxiliary power supply-3 detailed rules.

Its eight, safety certification-2 detailed rules.

Its nine, the load switch misaction detection function

Category 2 internal data / information / software formation and processing function (7 items and 24 detailed rules)

First, the data / information / software formation function

Second, the data / information / software processing function

Third, the rate and time period-4 detailed rules.

Fourth, zero-2 rules (electricity meter zero, demand zero).

Fifth, data storage-4 detailed rules.

Sixth, freezing-6 detailed rules (regular freezing, instantaneous freezing, daily freezing, agreed freezing, hour freezing, frozen content).

Its seven, show-8 detailed rules.

- -Category 2 extended function: (no).

-Comment on this article

The general data / information / software (multi-function meter, smart meter) output function of Q / GDW 13542013 standard summarizes the development and application experience of domestic such electricity meters in nearly 20 years, and complies with the new demand of metering technology in the development of smart grid in recent years. Its main characteristics:

Four types of electric energy metering are proposed, including active and reactive total power and active phase electric energy, combining active and reactive power, programming combination reactive power, time-sharing active and reactive electric energy.

The clock circuit and the timing method are proposed

Put forward 5 types of event records (22 rules): abnormal power supply work (6 rules), abnormal power grid operation (6 rules), electricity meter work record (5 rules), anti-theft function (3 rules), event reporting and recording (2 rules).

Clear 5 types of communication modes: RS485, infrared communication, carrier communication, public network communication, micro-power wireless communication.

Constant magnetic field monitoring and event recording are proposed as antipower theft measures.

The comparison function of electricity meter software is proposed

Put forward the error-action detection function of the load switch

However, there are some shortcomings in the standard preparation, especially:

First, the intelligence of smart meters: few automatic control and information interaction functions, inadequate requirements; two-way communication module requirements are not mentioned; no gateway application is introduced.

Second, it does not mention the metering base wave active electric energy and the harmonic active electric energy.

Third, it does not mention the requirements of IR 46 standard-active power meter international measurement proposal.

- -The standard is the centralized unit: the Science and Technology Department of the State Grid Corporation

The main drafters of the standard: Lin Fantao, Zhang Xin, Yang Xiangjiang and other 14 experts.

The following description describes the rise and evolution of multi-function meters, smart meters and the new generation of smart meters standards from 1995 to 2020.

2) In March 1995, the first domestic mechanical line standard JB / T 7656-95, multi-function Power meter was released.

First, the multi-function electrical meter is defined: a meter composed of measuring unit and data processing unit, which can complete a variety of billing methods, load management and long-distance data transmission functions.

Two is the meter type: according to the working principle into mechanical electronic multi-function meter; electronic multi-function meter.

Third, accuracy: active is 0.2,0.5 and 1.0; reactive is is 2.0.

Fourth, function:

-Basic functions of category 1 (12 and 30 rules)

Type 1 Data / Information / Software (General) output function (6 Item 25):

First, electric energy metering function: receiving the total active and (or) reactive power energy from the system; metering the total active and (or) total reactive power energy to the system;

Second, the measurement of peak, valley, peacetime section of active and (or) reactive power division-1 detailed rules.

Third, the measurement time interval is the maximum demand of 15min; the demand measurement mode can be interval or slip, with a slip interval of 1min.-3 Rules.

Fourth, with the external or internal attached programmer preset data.-7 Rules.

Fifth, the monitoring function-9 detailed rules.

Sixth, data transmission-5 detailed rules.

Category 2 internal data / information / software formation and processing function (6 items and 5 detailed rules):

First, the data / information / software formation function

Second, the data / information / software processing function

Third, the time period control function-3

detailed rules.

The fourth, show-2 detailed rules.

Fifth, the automatic reset zero function can also be manually reset zero.

Its six, have crash automatic recovery and program self-check function.

-Category 2 extended function (10 items)

It has the reading of active and reactive power during voltage transformer and fault checking, event recording and 10 functions of measuring V2 h and I2 h.

-Comment on this article

JB / T 7656-95 standard reflects the design level of domestic multi-functional electricity meters in the 1995s.

Type of meter: in 1995, induction meters continued to dominate the domestic meter market; electronic meters have been used in batches; multifunctional meters are in the low trial phase.

Class 3 multiplier:

First, the induction electricity meter uses a multiplier based on the principle of rotating magnetic field.

Second, the electronic electricity meter with the analog multiplier adopts the time-split multiplier.

Third, the electronic electricity meter with the digital multiplier usually uses the multiplier with the point product and the numerical integration algorithm.

The measurement error below 5% basic current is not assessed. Accuracy level, no 0.2S,0.5S level requirement.

Reactive power measurement, estimated to use only the sine wave 90 * phase shift circuit.

• multifunctional:

Compared with Q / GDW 13542013, JB / T 7656 / software-95 standard are rare: one rule of measurement type; not mentioned: active power combination, reactive power combination measurement, programming combination reactive power measurement, clock circuit and timing, event recording, anti-theft function, power-meter software, event recording, fault detection function, PT power recording and other important functions are included in the extended function.

Software: Preliminary requirements. Only propose the crash automatic recovery and program self-check function.

- -The drafting unit of the standard: Electric Power Department of the Ministry of Energy, East China Electric Power Joint Company and East China Branch of Electricity Consumption and Electricity Saving Committee of China Society of Electrical Engineering.

The main drafters of the standard: Huang Shouhai, Chen Li, Zeng Naihong and other 6 experts.

3) In May 1997, the power line standard DL / T 614-1997 "Multi-functional Electricity Meter" was released.

First, the definition of multifunctional electricity meter: composed of measurement unit and data processing unit, in addition to measuring active (reactive active) electric energy, but also has two or more functions such as time-sharing, measurement demand, and can display, store and output data of the electricity meter.

Second, the meter type: according to the working principle of the electromechanical multi-function meter; electronic multi-function meter

Third, accuracy: 0.2,0.5,1,2 and 3.

Fourth, function: (omitted)

-Comment on this article

DL / T 614-1997 standard on JB / T 7656-95 with some new technical requirements. But the release of the two standards is only two years apart, with few new requirements.

The requirement of functional function is put forward.

Accuracy: add 3 levels for reactive power measurement.

• function

First, type 1 data / information / software general output function: add event recording function-3 rules; not mentioned: active power combination, reactive power combination measurement, programming combination reactive power measurement, clock circuit and time calibration, 5 types of communication mode, anti-theft function, meter software comparison function.

Second, expansion function: add data communication function, daily load curve record, measurement of electric energy, loss, copper loss, load monitoring function, bar code number function.

- -The centralized unit: Technical Committee of Standardization of Electricity Measurement of the Ministry of Electric Power Industry.

The main drafters of the standard: Huang Shouhai, Wu Shijin, Yao Wenkui and other 5 experts.

4) In October 2007, China's first national standard GB / T 17215.301-2007 "Special Requirements for Multi-function Electricity Meter" was released

One is the multi-function watt-hour meter definition: by a single measurement mechanism, data processing unit, communication interface and other functional components and enclosed in a case, in addition to measuring and display active power, reactive power, also has time-sharing measurement, maximum demand measurement and so on two kinds or more functions, and storage and output data watt-hour meter.

Second, the type of meter: static meter

Third, accuracy: active power, 0.2S,0.5S, 1 and 2; reactive power, 2 and 3.

Fourth, function

-Basic functions of category 1 (12 Item 36)

Category 1 Data / Information / Software General output function (7 Item 20):

First, the measurement of unidirectional or bidirectional active electrical energy and total electrical energy, as well as the metering of bidirectional or four-quadrant reactive electrical energy.

Second, measure the active energy and reactive energy of each rate.

Third, the maximum demand measurement function-5 detailed rules.

Fourth, the setting of the clock

Fifth, the event recording function-9 detailed rules.

Sixth, measuring output device-4 detailed rules (pulse output sequence, photoelectric test output, work indicator, etc.).

Its seven, data communication-2 detailed rules (photoelectric interface, infrared communication, RS485 bus, etc.).

Category 2 internal data / information / software formation and processing function (5 items and 16 detailed rules)

First, the data / information formation function

Second, the data / information processing function

Third, the rate and time period setting function-3 detailed rules.

Fourth, the measurement of the data storage function-4 detailed rules.

Fifth, show-9 detailed rules.

-Category 2 extended function (9): prepaid; calculating 11 functions including apparent power, loss, copper loss or other required parameters.

-Comment on this article

In 2005, the total output of domestic electronic meters surpassed induction meters for the first time; later, electronic meters dominated the domestic meter market. From the national standard GB / T 17215.301-2007 release, we can see the domestic multi-functional table technology for 10 years of development level.

The multifunction table is defined more comprehensively

Accuracy: Level 0.2S,0.5S. Measurement measurement assessment from 1% basic current.

• function

General output function of data / information / software: adding event recording rules (9 articles); proposing data communication technical requirements and communication protocol; not mentioned: active power combination, reactive power combination measurement, programming combination, clock circuit requirements, 3 communication modes (carrier communication, public network communication, micro power wireless communication), anti-magnetic field interference, and comparison function of electricity meter software.

Extended feature: Represent the prepaid feature.

- -The standard centralized unit: National Electrical Instrument Standardization Technical

Committee.

The main drafters of the standard: Xu Min, Wang Zhaohong, Xu Heping and 17 experts.

5) In December 2007, the power line standard DL / T 614-2007 "Multi-functional Electricity Meter" was released

First, the multi-function table definition: the same as DL / T 6141997.

Second, the type of meter: static meter

The third is accuracy: divided into four grades: 0.2S,0.5S, 1 and 2.

Fourth, function

- -Basic functions of category 1 (16 items and 51 detailed rules)

Category 1 Data / Information / Software General output function (8 Item 28):

First, metering one-way or two-way active electric energy, one-way or four-quadrant reactive electric energy.

Second, the measurement of multiple period of active electric energy, reactive electric energy.

Third, the demand measurement-6 detailed rules.

Fourth, the event records-9 detailed rules.

Fifth, the type of communication interface-2 detailed rules (communication interface, communication requirements, communication protocol).

Sixth, the communication interface requirements-6 detailed rules.

Its seven, pulse output-3 detailed rules.

Eight, the power grid operation parameter measurement-2 detailed rules.

Category 2 internal data / information / software formation and processing functions (8 items and 23 detailed rules):

First, the data / information / software formation function

Second, the data / information / software processing function

Third, the calendar, the rate, the time period-2 detailed rules.

Fourth, the measurement of the data storage function-3 detailed rules.

Fifth, freeze-3 detailed rules.

Its six, clear zero out-2 detailed rules.

Its seven, show-8 detailed rules.

Eight, software requirements-5 rules (including: manufacturers to provide operation application software, software is not allowed after the factory remote and on-site upgrade update, etc.).

- -Category 2 extended function (5 items): metering and visual electric energy; harmonic voltage and current monitoring; calculation loss and copper loss; optical fiber, Bluetooth and wireless communication mode.

-Comment on this article

Compared with the previously released multi-functional table mechanical lines, power lines and national standards, the technical requirements of the multi-functional table of DL / T 614-2007 standard can basically cover:

First, general functions of data / information / software: electric energy measurement, demand measurement, event recording, communication interface and communication requirements, grid operation parameter monitoring, pulse output; data / information / software formation, processing functions: rate, data storage, freezing, zero clearance, display are relatively complete, and electric meter software is not allowed to upgrade after delivery; but not mentioned: active power combination, reactive power combination measurement, programming combination reactive power measurement, clock circuit requirements, three types of communication (carrier communication, public network communication, micro power wireless communication), anti-magnetic field interference power stealing function, and comparison function of electricity meter software.

Second, expand the function: mention harmonic voltage, current monitoring; optical fiber, Bluetooth, wireless communication applications.

6) In September 2009, China's first state Grid standard Q / GDW 354-2009 "Smart Power Memeter Functional Specification" was released.

First, the definition of smart electricity meter: the same as the previous state grid standard Q / GDW 1354-2013 "Smart electricity Meter Function

Specification".

Second, the type of meter: static meter

Third, accuracy: 0.2S,0.5S, 1,2 four levels.

Fourth, function

- -Basic functions of category 1 (25 items and 85 detailed rules)

Data / information / software output function (18 items and 65 detailed rules)

The following is the general (multi-function meter, smart meter) output function (11 items and 40 detailed rules):

First, forward and reverse total active power and phased active power and four quadrant total reactive power measurement-1 rules (do not use the phased active power arithmetic addition method to calculate the total active power).

Second, the combined active power and the combined reactive power electric energy metering.

Third, time-sharing active electric energy, reactive electric energy measurement-2 detailed rules.

Fourth, the demand measurement-4 detailed rules.

Fifth, the clock circuit and the school time-5 detailed rules.

Its six, the event record-15 detailed rules.

Its seven, communication requirements-2 detailed rules.

Eight, communication mode type-4 detailed rules (RS485, infrared communication, carrier communication, public network communication).

Its nine, signal output-3 detailed rules (pulse output, multi-function signal output, control output).

Its ten, power grid operation parameters monitoring-2 detailed rules.

Its 11, security protection-2 detailed rules (programming, reading table operation).

The following is the special (smart meter) output function (7 items and 25 detailed rules):

First, the fee control function-11 detailed rules.

Second, the load record-3 detailed rules.

Third, the ladder of electricity price

Fourth, power failure meter reading-2 detailed rules.

Fifth, the alarm-4 detailed rules.

Sixth, the auxiliary power supply-3 detailed rules.

Its seven, safety certification-2 detailed rules.

Internal data / information / software formation and processing functions (7 items and 20 detailed rules):

First, the data / information / software formation function

Second, the data / information / software processing function

Third, the rate and time period-4 detailed rules.

Fourth, zero-2 rules (electricity meter zero, demand zero).

Fifth, data storage-2 detailed rules.

Its six, freeze-5 rules (regular zero, instantaneous zero, daily zero, agreed zero, hour zero).

Its seven, show-7 detailed rules.

- -Category 2 extended function: (no).

-Comment on this article

The development and application from domestic multi-functional meter to smart electricity meter is a new leap in the development of domestic electricity meter technology.

State Grid launched Q / GDW354-2009 "Functional Specification for Smart Electricity Meters", which plays a connecting role in the evolution of China's multi-function meter standards in China.

For the previous power standard DL / T 614 / 2007 standard, Q / GDW 354-2009 standard is advanced, proposing the active power combination, clock circuit measurement, increasing event recording rules, carrier communication, public network communication mode; rate function, load record, ladder electricity price, power failure meter reading, safety authentication function, etc.

For the subsequent Q / GDW 1354-2013 standard, Q / GDW 2009 standard is late, not mentioned: programming combination reactive power metering, micro power wireless communication, anti-magnetic field interference theft function, meter software comparison function, load switch misaction detection, intelligent items, base wave active energy, harmonic active energy metering.

7) In August 2020, the State Grid Standard General Technical Specification for Smart Power Meters (2020 Edition) IV. Functional Specification for Smart Electricity Meters was released.

State Grid Smart meter (2020 edition) is a new generation of smart meters. The four specifications of the General Technical Specification for Smart Electricity Meters (2020 Edition) respectively quote the relevant requirements of IR 46 international measurement proposal of active power meters.

First, the definition of smart electricity meter: the same as Q / GDW 1354-2013 Functional Specification of Smart Electricity Meter.

Second, the type of meter: static meter

Third, accuracy: active power measurement, B, C, D; reactive power measurement, 2.

Fourth, function

- -Category 1, basic functions

· Category 1 data / information / software output function:

First, it is basically the same as Q / GDW 1354-2013 standard (12 items): measuring total active and reactive power, combined power, time-sharing power, phase separation power, demand measurement, clock and timing, communication requirements, communication mode type, signal output, constant magnetic field monitoring, power meter software comparison function.

Different functions:

Second, write down 26 detailed rules for the event record, and add 4 detailed rules (clock fault record, metering chip fault record, electricity meter change current abnormal record, etc.).

Third, the power grid operation parameters monitoring-7 rules, add 3 rules (three-phase electricity meter to provide beyond the limit monitoring function, etc.).

The following is the dedicated (smart meter) output function:

First, it is basically the same as Q / GDW 1354-2013 standard (7 items): cost control function, ladder electricity price, power failure meter reading, power protection function, alarm, auxiliary power supply, load switch misoperation detection.

Second, delete 2 items: load record, safety certification.

· Category 2 internal data / information / software formation, processing functions, with some changes.

First, it is basically the same as Q / GDW 1354-2013 standard (4 items): data / information / software formation, processing function, rate and time period, and display function.

Different functions:

Second, zero-3 rules, add 1 rules (event zero).

Third, freeze-10 rules, add 4 rules (minute freeze, monthly freeze, settlement day freeze, ladder settlement freeze).

Fourth, delete the function: data storage

- -Category 2 extended function: (no).

-Comment on this article

The 2020 version of the functions of Q / GDW 1354-2013 standard: 12 data / information / software general output functions, 7 special output functions, 4 internal data / information / software formation and processing functions; 4 functions added; 3 functions deleted. Note that Q / GDW 1354-2013 "Smart Electricity Meter Function Specification", after 7 years of application, there are still 76% of the functions are applicable.

The biggest change in the function specification of the 2020 smart meter is accuracy: active power measurement, grade B, C, D; reactive power measurement, and level 2. Here, the representation method of IR 46 international measurement method of active energy meter still adopts the representation method of IEC meter standard. This complicates the detection of active and reactive power.

At the same time, no new functions are proposed for the development and application of fundamental and harmonic active power metering of intelligent projects of electricity meters.

8) About the latest standard: "Intelligent Internet of Things Energy Meter Function and Software Specification"

In January 2022, State Grid issued Q / GDW 12180-2021 "Specification for Intelligent Internet of Things Power Meter Function and Software".

However, 2022 is the second year of the bidding of the first batch, which only accounts for 2.1% of the total equipment of the same batch. Meanwhile, there are some disputes about the application of intelligent meter, mainly high meter price; extended function is optional; the first use of embedded general operating system is not practical, and some special requirements of electricity meter require the development of embedded special operating system.

In view of the above situation, this article will temporarily not discuss the state Grid Q / GDW 12180-2021 "Intelligent Internet of Things Energy Meter Function and Software Specification".

## 4. The origin and evolution of domestic multi-functional electricity meter products

Domestic multi-function meter: from 1992 to 2021, 30 years of development, from the mechanical and electronic multi-function meter, through the use of analog multiplier multi-function meter, with digital multiplier (four types) multi-function meter, smart meter to the new generation of smart electricity meter 5 generation product technology evolution.

The evaluation of the technical performance of multi-function table mainly has four aspects: multi-function table structure, new measurement technology; accuracy; multi-function / intelligence.

1) In 1992, Shanghai Electric degree meter Factory first launched the independently designed DSD 8,0.5 level, mechanical and electronic three-phase three-line multi-function meter.

- -Multi-function meter structure: it is composed of induction base meter, infrared photoelectric sensor and electronic multi-function recorder.

- -Using a power multiplier based on the principle of a rotating magnetic field.

-Multi-function: total active power metering; time-sharing energy metering: maximum demand of interval or slip type; magnetic suction infrared communication; self-test function; strong anti-interference ability.

-Comment on this article

Mechanical and electronic multi-function meter is the first generation of multi-function meter in China, which realizes the multi-function based on the measurement of active power and electric energy. Because there is no basic technology of reactive power measurement, the development of multifunction is limited.

2) In August 1994, Weisheng first developed the DSSD331-type / DTSD341-type and 0.5S class electronic three-phase three-line / three-phase four-line multifunction table using digital multiplier

- -Multi-function table structure

: All-electronic multi-function meter is composed of current sensor (CT), voltage sensor (PT), analog-to-digital converter (A / D), microcomputer special chip and display (LED) and other parts.

- -Digital multiplier: in the microcomputer special chip, the digital current, voltage instantaneous value of various judgment and processing and operation.

— multifunctional

Forward and reverse active power, perceptual, capacitive reactive power and four quadrant reactive power.

Time-sharing metering of active electric energy and reactive electric energy.

Maximum demand for active power and reactive power.

Power grid active power, reactive power, voltage, current, power factor and frequency monitoring.

Load curve record

Two independent simultaneous RS485 interface, adsorption infrared interface.

4-way empty contact pulse auxiliary terminal output

Event record of 8 classes

Self-diagnosis function and fault alarm function

Voltage and current harmonic analysis function, up to 21 harmonics.

- -The new electricity meter has passed the ministerial appraisal

In the same year, the new electricity meter passed the ministerial appraisal under the auspices of the basic Equipment Department of the former Machinery Department.

Director of the appraisal Committee: Professor Fei Zhengsheng

The expert that the electric power department attends the appraisal meeting has Qu Tao, Chen Li, Zong Jianhua.

Conclusion: the performance index and appearance structure of the product reach the advanced level of similar products in the early 1990s, and are in the leading position of similar products in China.

-Comment on this article

Weisheng is the first and the first to launch the multi-function table using digital multiplier.

The multi-function table of the digital multiplier is the first class of the third generation of the domestic microcomputer chip. Through the sampling of periodic waveforms, multi-functional operation can be carried out, including power grid voltage, current, phase angle, active power, reactive power, apparent power, frequency, harmonic content and so on.

At first, the digital multiplier applies only to low-level multifunctional tables. Later, by improving the sampling rate and calculation accuracy, it can be applied to the multi-function meter and the standard electricity meter at the 0.1S class power grid pass.

Now, the use of digital multiplier multi-function table, with easy to achieve multi-function, low cost, short product development cycle advantages, has become the mainstream of domestic multi-function table.

However, the A / D sampling software of the digital multiplier is non-real-time and has great drawbacks. For aperiodic and unstable signals and for scenarios with high harmonic content, the active power cannot be accurately calculated. At the same time, even if the hardware of different manufacturers' products is the same, the actual quality and performance level may also be quite different, mainly because the software capability of manufacturers can not be assessed by a unified standard.

3) In December 1995, Ning Photoelectric Factory launched the DSSF22A-X / DTSF22A-X, 1-level electronic three-phase three-line / three-phase four-stage intelligent multi-function table with the integrated circuit.

- -Multi-function table structure: The multi-function table adopts special integrated circuit, non-volatile memory, special electric card, large screen liquid crystal display and other technologies.

- -Analog multiplier: It is understood that in 1991, Ning Photoelectric factory commissioned 771 Institute of China Aerospace Academy to design small-scale metering special integrated circuit, using the principle of time-split multiplier and double integral V / F conversion.

— multifunctional

Accurate measurement of active power, reactive power power, active power demand, reactive power demand, power factor, prepaid payment and 4 rates.

Overload and overpreset power alarm and automatic power off.

Abof phase alarm record

Autopsy meter reading

The operation of card insertion and card pulling realizes the information transmission between the user's electricity meter and the electricity sales management system of the power supply department.

Adopt a variety of software, hardware anti-interference measures.

-Comment on this article

Ning photoelectric factory: launched the electronic multi-function table using the divided multiplier integrated circuit earlier in China.

The multi-function table using the analog multiplier is the second generation of domestic multi-function table.

The time split multiplier plays a very important role in the active power metering technology. The (D) C12 standard power converter with 0.005% measurement accuracy is the standard electricity meter for the distortion of the voltage and current signals, the harmonic power measurement value of the split multiplier is consistent with the theoretical calculation result.

In 1973, Jiangxi Electric Power Experimental Research Institute developed the PS-4 single-phase electronic standard electricity meter for the first time in China, using the principle of time-split multiplier.

However, the time-split multiplier is only used in active power measurement. As a multi-function table, its reactive power measurement and other functions are important components, which need to be supplemented with alternative phase shift circuit or A / D sampling scheme to produce reactive power measurement and other functions, which is complicated. Therefore, at present, the market share of the multifunctional meter is not large.

4) In 2002, Nanjing Bluestar Electric Power Instrument Research Institute launched the 0.5S class, full digital stationary multi-function meter.

-Multi-function table structure: composed of full digital CS5460A-type measurement chip (integrated A / D, voltage reference, operation and discharge, DSP and digital interface), digital signal operation, digital filtering, digital quantity correction, digital adjustment unit, etc., without any physical adjustment element.

- -Digital multiplier: The CS5460A chip adopts a new generation of analog-to-digital conversion technology and a DSP digital signal processor.

— multifunctional

Active electric power and reactive electric energy metering.

Phase electric energy metering

Voltage, current effective value, instantaneous value, power, power factor.

Full digital compensation, including phase compensation, without external devices.

Realize the treatment of the measured cross variable range, and make nonlinear compensation for the wider measurement range and the transformer.

Clock backup super lithium battery with a storage life of more than 20 years.

-Comment on this article

Nanjing Blue Star launched the full digital multi-functional table, is the third generation of the second type of digital metering chip multi-functional table. Its characteristics of full numbers: measurement accuracy, anti-interference, reliability quality have been greatly improved.

Nanjing Bluestar, founded around 1980, is mainly engaged in electric power instrument, instrument data acquisition device and software development, marketing and import and export business.

It is estimated that from 2002 to 2008, Bluestar's full digital electronic multi-functional meter labeling products sold well, mainly supplying the short-term OEM demand of multi-functional meter for the transformation of large induction meter enterprises, and the short-term labeling needs of emerging electronic meter enterprises with insufficient multi-functional meter development capacity.

5) In 2002, Weisheng developed a 0.2S/0.5S-class electronic multifunctional meter with harmonic metering function for the first time in China

This part is extracted from the manual of DSSD331 / DTSD341-9A / B / C electronic multi-function electricity meter.

- -Multi-function meter structure: adopt the international popular high-end electricity meter design scheme: metering DSP + management MCU hardware scheme.

- -Harmonic analysis and measurement principle: The watt-hour meter is sampled and digitally processed by 16-bit A / D and DSP high-speed processors for the current and voltage of each phase. Through the corresponding digital calculation, that is, DSP transforms the current and voltage signal with 256 points FFT transformation in real time, which can give the current, voltage and harmonic components 2-49 times. The harmonic amplitude accuracy is 2%, and the phase accuracy is + / -1-2 *. DSP completes the electrical parameter measurement, electric energy accumulation, harmonic analysis, base wave electric power, and harmonic electric energy calculation; metering data is exchanged with management MCU; management part adopts display, data statistics, storage, communication, meter function selection, and initialization data setting.

— multifunctional

Total active-power and time-of-service electricity.

Metering the combined reactive electric energy

Vacation settlement across the month

Maximum demand amount calculation

Measure the voltage, current, and power

Harmonic electric energy metering:

Harmonic metering mode (3 kinds) is optional:

The first is based on base wave measurement:

total active electric energy = base wave active electric energy. Its active power energy metering is 0.5S class.

The second measurement method based on digital multiplier: total active electric power = base wave active electric energy + total amount of harmonic active electric energy. Its active power energy metering is 0.2S class.

The third harmonic measurement method based on Fourier transform: total active power = base wave active power + the absolute value of the total amount of harmonic active energy. Its active power measurement is of 0.5S class.

· Transformer loss compensation

· Correction and compensation of external PT, CT ratio difference and angular difference.

· Communication interface: 2 independent RS485 interfaces, 1 adsorbed infrared and far-infrared interface.

· load record

· incident record

· Test output

· Alarm function

· Self-diagnosis and recovery function of the electricity meter

-Comment on this article

Weisheng for the first time in China launched 0.2S/0.5S level, with harmonic metering function of all electronic multi-function table, is the third generation of domestic third class of (measurement DSP + management MCU) high-end scheme of multi-function table.

Application prospect of different measurement methods of the new multi-function table:

· The second measurement mode based on digital multiplier, active power measurement level 0.2S, is adapted to the billing application under the current electricity price policy.

· The first is based on base wave measurement, active power measurement 0.5S. At present, Canada has officially implemented the-wave metering. And domestic, need to strive for the national electricity price New Deal to apply. At the same time, it is necessary to study the active energy measurement after FFT conversion to 0.2S.

· The third harmonic measurement mode based on FFT, the active power measurement is 0.5S level. At present, the electricity price administrative department has not yet issued to the harmonic source users penalty, this measurement method is difficult to implement temporarily.

6) In 2011, Weisheng first launched the DTSD341-MA1V1.0,0.1S level high accuracy settlement pass multi-function table in China.

This part is extracted from Weisheng "High-end metrology Weisheng Pilot", and edited.

- -Multi-function table structure:

· Pass table: mainly by I / V conversion circuit (CT), V / V conversion circuit (resistance partial voltage), respectively, to the analog-to-digital converter (A / D) -with independent voltage reference, and then to the metering DSP.

· Key components selected international high reliable brand components, such as German VAC high permeability current transformer, the VISHAY high precision column resistance, the American 18 road 24 high precision mode digital converter (A/D)-12.8k sampling rate, the ADI high performance external voltage reference chip and high speed DSP metering chip, using analog separation multilayer high performance PCB, the overall measurement scheme has high accuracy, high stability, high reliability, long life characteristics.

- -High-precision design

· Apply NewtonCotes high-price integration algorithm, fast power and power current resolution technology, and hardware phase matching technology to achieve high accuracy measurement under dynamic load.

· The accuracy of the factory active power metering is controlled at less than + / -0.04%; the reactive power metering is of grade 0.5S.

· The impact of ambient temperature, power factor, frequency, and load current fluctuation and variation on measurement accuracy can be ignored.

· The measurement accuracy is less than 0.1%.

— multifunctional

· Forward active electric energy and four-quadrant reactive electric energy metering

(reverse metering is for reference only).

· Two combinations of reactive power metering, which can be arbitrarily combined by four-quadrant reactive electric power.

· Time-sharing active power quantity and reactive power quantity.

· Calculate the maximum demand of active power and reactive power by time-sharing.

· Up to 12 rates, two sets of main and secondary periods, clock double backup.

· 6 types of data load curve records can be set, with a capacity of 20M bytes.

· Dual RS485 interface; adsorption / far infrared communication interface; 100M adaptive Ethernet interface, support UDP / TCP remote meter reading (DL / T 645-2007 or user custom protocol).

· Rich in power grid event records

· Double-backup data storage, with self-inspection and error correction functions.

· 4-way empty contact power pulse and LED power pulse output

· Large screen with wide view LCD display.

— product identification

In December 2010, the scientific and technological achievement appraisal meeting of "Research and development of 0.1S level key technology of power meter" organized by Hunan Science and Technology Department was held in Weisheng Science and Technology Park.

Appraisal opinion: "This achievement has a number of independent intellectual property rights, and the key technology has reached the international advanced level. It is suggested to develop finalized products and realize industrialization as soon as possible".

-Comment on this article

Weisheng launched 0.1S class, high accuracy settlement threshold multi-function meter, is the third generation of domestic multi-functional meter using high-end metering algorithm, filling the gap of domestic high-end electricity meter products, exceeding the highest level of 2011 IEC electricity meter standard 0.2S.

At present, the 0.1S

The grade pass table has been in stable and reliable operation for more than 10 years, at the regional power grid, provincial power grid, power plant and the electrified railway with frequent load changes. At the same time, the 0.1S level pass table has passed the EU MID certification and DLMS certification, and has been applied in many countries.

In 2018, Langier released in the Shandong Institute of Electricity Research Institute: Langier is developing a new generation of E860 series higher precision settlement threshold table. The development of E860 series 3-core electricity meter, its accuracy meets the 0.1S level requirements of IEC standard, and has more advanced anti-theft capability in software and hardware; it is understood that in the mid-2023, Langier will release E860 series and 0.1S level meter to the international market. The domestic electricity meter industry should pay special attention to the secondary information, focusing on the factory control error, the reverse metering accuracy, the requirements of the base wave and harmonic active power metering, etc.

7) In 2013, State Grid launched the 2013 version of state Grid standard smart electricity meter

This smart meter is the second smart meter after the 2009 standard smart meter of state Grid. The design and production of smart meter standards, grid access testing, product bidding and acceptance are provided by grid metering and material departments.

- -Structure and design of technology of smart meters

· Structural design: According to Q / GDW 1356-2013 Specification for Three-Phase Intelligent Power Meters and Q / GDW 1355-2013 Specification for Single-Phase Intelligent Power Meters, the structural design of smart meters includes:

First, the electricity meter parameters are determined

Second, display, indicator light, power failure display

Third, the appearance structure and the installation size

Fourth, the material and process requirements

· Technical design of smart electricity meters:

According to Q / GDW 1827-2013 Technical Specification for Three-phase Intelligent Electricity Meters and Q / GDW 1364-2013 Technical Specifications for Single-phase Intelligent Electricity Meters, the technical design of smart electricity meters, including:

    First, the specification requirements,

    Second, environmental conditions

    Third, mechanical and structural requirements

    Fourth, the accuracy requirements

    Fifth, the electrical requirements

    Its six, the insulation performance

    Seventh, the electromagnetic compatibility requirements

    Its eight, reliability requirements

    Ninth, data security requirements

    Its ten, the software requirements

    Its eleven, packaging requirements

Twelve, the communication module interchangeability requirements.

- -Special chip for electric energy metering

At present, there are four types of special chips for energy metering: three-phase energy metering chip, three-phase multi-function metering chip, single-phase energy metering chip, single-phase energy meter SOC chip.

In China, Shanghai Juquan, Shanghai Beilin, Fudan Microelectronics, Shenzhen Xinhai, occupy most of the domestic electricity meter market; foreign manufacturers ADI, TDK, Atemel, Cirrus Logic, occupy part of the domestic electricity meter market.

Typical case of special chip for electricity meter:

· Shanghai Bell: BL6513C three-phase active power metering chip based on digital signal processor with input dynamic range of 3000:1; nonlinear error is less than 0.1%; static power consumption of chip is 25 mW; measure forward and reverse active power.

· Shanghai Juquan: ATT7022E / 26E / 28E series high-precision, multi-function three-phase electric energy metering chip based on digital signal processor, integrated with multi-channel second-order Sigma-Delta ADC, reference voltage, digital signal processing circuit, and built-in temperature sensor. The input dynamic range is 5,000:1, and the nonlinear error is less than 0.1%. Measure the active power, reactive power, apparent power, active power and reactive power of each phase, and can also measure the current, voltage, effective value, phase Angle, power factor, frequency and other parameters of each phase. Can accurately measure 41 harmonic active, reactive, apparent power.

· ADI company: 4 high-precision energy metering IC

ADI's ADE7878, ADE7868, ADE7858, ADE7854 high precision electric energy metering IC can provide 0.2% accuracy of active and reactive power measurement in the dynamic range of 3000:1. At the same time, the reactive power and active power can also be measured with an accuracy of 0.1%. One of the chips has a fundamental wave electric energy measurement function.

— multifunctional

According to the requirements of Q / GDW 1354-2013 "Functional Specification for Smart Power Meters" and Q / GDW 1365-2013 "Technical Specification for Security Certification of Smart Power Meters", the 2013 version of state Grid standard smart meter has relatively complete functions. Its functions: (omitted).

-Comment on this article

State Grid 2013 version of the standard smart meter is the third generation of domestic second-type multifunctional meter using digital metering chips.

After the 2009 version of the standard smart meter of the state grid operated for 1-3 years, the meter failed many, reflecting serious product quality problems. To this end, the Marketing Department of State Grid issued the Notice on rectifying and Strengthening the Quality Management of smart meters, and organized the formulation of technical specifications for the main components of smart meters (13 items). Subsequently, the 2013 version of smart meter standard.

2013 version of the standard smart meter products add new requirements: constant magnetic field protection, power frequency magnetic field potential, high temperature limit work, communication module interface with load capacity, communication module interchangeability, meter software comparison function, improve the smart meter measurement stability, reliability, power theft ability, communication module connectivity ability, and meter failure rate greatly reduced. However, no new functions are put forward for the development and application of fundamental and harmonic active power metering of intelligent electric meters.

8) In 2020, State Grid launched the 2020 version of state Grid standard smart meter- -a new generation of smart meters.

The 2020 version of the state grid standard smart meter is developed on the basis of the 2013 version of the state grid standard smart meter, citing the key requirements of IR 46 active power meter international measurement proposal.

- -Structural design of smart electricity meters

According to its standard " three-phase intelligent watt-hour meter (2020 edition) general technical specification: three, three-phase intelligent electricity meter type specification, the single phase intelligent watt-hour meter (2020 edition) general technical specification: three, single phase intelligent electricity meter type specification requirements, 2020 version of smart meter structure design, including specifications, display, indicator, power outage display, shape structure and installation size, materials and process requirements. Among them, the current specification is expressed, using the minimum current Imin, the turning current Itr, and the maximum current Imax.

— technical design

In accordance with the requirements of the State Grid standard "General Technical Specification for Three-Phase Intelligent Electricity Meters (2020 Edition): II, Technical Specification for Three-Phase Intelligent Electricity Meters" and "General Technical Specification for Single Phase Intelligent Electricity Meters (2020 Edition): II. Technical Specification for Single Phase Intelligent Electricity Meter", the technical design of smart meters in the 2020 edition, including:

· Specification requirements

First, accuracy level: active power measurement, A, B, C, D; reactive power measurement, 2.

Second, current specification: the minimum current Imin, turning current Itr, maximum current Imax.

· ambient condition

· Mechanical and structural requirements

· Accuracy requirements

First, the factory error of electricity meters is expressed by the basic maximum allowable error.

Second, timing accuracy, add a detailed rules requirements.

Third, the impact of the quantity, the reference of IR 46 standard requirements.

Harmonic test: harmonic-square top wave waveform test in voltage and current circuit; harmonic-spike wave waveform test in voltage and current circuit; harmonic / pulse string triggered waveform test in current circuit; odd harmonic-90 degrees phase-triggered-waveform test in current circuit.

Interference impact test: citing multiple electromagnetic compatibility test items: (omitted).

· Electrical performance requirements

· insulating property

· Reliability requirements

· Data security requirements

· Software requirements

· packing requirement

· Interchangeability requirements for communication modules.

- -IR 46 International measurement proposal of active power meter and IEC meter standard

· Accuracy representation method: maximum allowable error, basic maximum allowable error, and maximum allowable error deviation.

· Traceability of errors: The data change process of the electricity meter from the initial inherent error, the inherent error, the anomaly (falut) to the major anomaly (falut) should be traceable.

· Accuracy level: A, B, C, D, which covers A wide range of changes, including the reference amount, influence amount and interference amount.

· Current working range representation of electricity meter: starting current Ist, minimum current

Imin, turning current Itr, and maximum current Imax. Specifically, there are accuracy calculation methods from Ist to Imin.

· Accuracy composition: including the basic maximum allowable error, the allowable error of the influence amount, and the allowable influence of the interference. For the first time, the limit of interference amount, interference level, allowable influence and electricity meter error deviation is proposed.

· Combined error estimation

· Metering characteristic protection- -hardware separation and software separation of electricity meters.

— multifunctional

According to the functional specification requirements of state Grid smart meter (2020 version), the 2020 version of smart meter products have more complete functions. Its functions: (omitted).

-Comment on this article

State Grid 2020 version of the standard smart meter is the third generation of the second type of domestic multi-function meters using digital metering chips.

The main feature of the 2020 version of the standard smart meter is the application and demonstration of IR 46 standard on domestic smart meters for the first time, focusing on the measurement accuracy and measurement performance protection requirements of IR 46.

In recent years, IR 46 standard has become an important content and basis for the formulation of smart electricity meter standards and national metrological verification regulations for electricity meters. However, IR 46 standard has limitations. At present, only the international measurement proposal of active power meters. The reactive power measurement test of multi-function meters needs to follow the IEC meter standard, which is complicated.

**epilogue**

Reviewing the emergence and evolution history of domestic multi-functional meter (smart meters, new generation smart meters), we can see that after 30 years of development, the domestic multi-functional meter industry has undergone profound changes:

· Domestic multi-functional meter type: from the mechanical and electronic multi-functional meter in 1992, through the electronic multi-functional meter and smart electricity meter, to the new generation of smart electricity meter in 2020.

· Design function of multi-function meter: from JB / T7656-95 multi-function electricity meter standard has 12 basic functions and 30 rules, to Q / GDW 1354-2013 "Smart Power Meter Function Specification" has 30 basic functions and 119 rules.

· Accuracy of multi-function table: developed from active power metering level 1 in 1992 to level 0.1S in 2010 (using digital multiplier).

· The 2020 version of state Grid smart meter first cited the key requirements of IR 46 active power meter on the domestic electricity meter.

Reviewing the development process of the above domestic multi-functional meter, the electricity meter industry should bear in mind the achievements of the following units in the development exploration and guidance of domestic multi-functional meter:

· East China Electric Power Administration: In 1995, it issued the first electric Power Measurement Modernization-Current Report of Power Measurement Technology of East China Power Grid, and organized the drafting of the mechanical and power standards of the first multi-functional table.

· Jiangxi Electric Power Experimental Institute: the first electronic standard electricity meter was developed in China in 1973; later, Ning Optoelectronics Plant introduced the electronic multi-function meter with special integrated circuit for the first time in China.

· Wei Sheng:

In 1994, the multi-function table using digital multiplier (microcomputer chip) was first launched in China.

In 2002, the first 0.2S class electronic multi-function table with harmonic metering function was developed.

In 2010, the 0.1S class high accuracy settlement pass electricity meter was launched for the first time in China.

· In 2013, led by China Electric Power Research Institute, Weisheng, Linyang and other electricity meter enterprises jointly launched the 2013 version of State Grid standard smart electricity meter.

In 2020, state Grid will launch a new generation of smart meters-smart meters (2020 version).

**reference documentation**

[1] Wang Changqing (supervisor: Tong Weiming) Research on smart grid advanced measuring system smart electricity meter "Master thesis of Harbin Institute of Technology" -2011-06-01

[2] Wang Jingzhe lost pressure loss timer does not need to configure several conditions "Industry and Technology Forum" -2011-05-30

[3] Li Ming; Li Chunzhang, Zhu Yang-2010-07-18

[4] Lu Ying (supervisor: Wu Guozhong) Design research and application of MSP430 electricity meter "Master's thesis of Zhejiang University" -2010-05-01

[5] Yue Yaobin; Meng Xiangzhong; Design of Microcomputer Application-2006-07-15

[6] Yue Hu; Guo Wanzhu; Liu Zhuo; Industrial Measurement-2011-09-15